%% file: paper.tex
\documentclass[12pt]{iopart}

\usepackage{graphicx}
\usepackage{color}
\usepackage{epsfig}
\usepackage{float}
\usepackage{multirow}
\expandafter\let\csname equation*\endcsname\relax
\expandafter\let\csname endequation*\endcsname\relax
\usepackage{amsfonts}
\usepackage{amssymb}
\usepackage{amsmath}
\usepackage{tabularx,url,color}
\usepackage{hyperref}

\usepackage{lineno}
\usepackage{upgreek}
\usepackage{comment}

\begin{document}
\leftline{Dated: \today}

\title{First Tests of a Newtonian Calibrator on an Interferometric Gravitational Wave Detector}
\author{D. Estevez, B. Lieunard, F. Marion, B. Mours, L. Rolland, D.Verkindt}
\address{Laboratoire d’Annecy de Physique des Particules (LAPP), Univ.  Grenoble Alpes,
Universit\'e Savoie Mont Blanc, CNRS/IN2P3, F-74941 Annecy, France}
mailto{ mours@lapp.in2p3.fr}
\date{\today}

\begin{abstract}
The ongoing improvements of the advanced gravitational wave (GW) detectors are setting challenging requirements 
on instrument calibration. 
We report tests of a calibration technique, based on the well-known gravitational force, 
which has been applied for the first time on a large interferometer. 
The results obtained with Advanced Virgo are in good agreement with the predictions 
and with the usual calibration method. 
This technique is expected to lead to accurate absolute calibration at the sub-percent level
in the coming years, matching the needs of the rapidly evolving GW science.
\end{abstract}

\maketitle


\input{intro}

\input{model}

\input{proto}

\input{test}

\input{coupling}

\input{next}

\input{conclusion}

\section*{Acknowledgements}
We are indebted to the Virgo Collaboration for allowing us to use the data
collected during the tests reported here, and are grateful for support
provided by the Virgo Collaboration and the European Gravitational
Observatory during those tests. We thank our colleagues in the
Virgo Collaboration and in the LIGO Scientific Collaboration for useful
discussions. We thank the technical staff at LAPP for their help in
building the NCal prototype, especially L. Giacobone for machining parts
and S. Petit for contributing to the device controls.

\section*{References}

\bibliographystyle{iopart-num}
\bibliography{references}

\end{document}

%% file: intro.tex
\section{Introduction}
The advanced LIGO and Virgo detectors~\cite{LIGO,Virgo} 
have started to observe sources of gravitational waves~\cite{GW150914,GW170814,GW170817}.
The quality of their observations depends on their accurate calibration,
both relative calibration for example for sky localization of the source,
and absolute calibration for example to determine the Hubble constant~\cite{GW170817Hubble}.
The planned improvement of the network sensitivity~\cite{ObservingScenarios}
is going to set even more stringent requirements on the calibration of the instruments.

Calibrating the detectors needs an absolute reference.
Up to now, this has been done relying either on the laser wavelength,
through the "free Michelson" technique~\cite{calibVirgo2011}, or on the power of an auxiliary laser beam, 
using so-called “photon calibrators”~\cite{calibLIGO,calibVirgo2018}.
Another option is to use the variation of the Newtonian gravitational field produced 
by moving masses to induce a known displacement of the interferometer mirrors.
This is an old idea derived from the Cavendish experiment and motivated by tests of the gravitation law, 
explored with resonant bar detectors~\cite{Sinsky,Hirakawa80,Kuroda85,Mio87,Explorer91,Explorer98}, 
already proposed at the time of the Virgo conceptual design~\cite{FCD},
discussed by multiple authors~\cite{Hild07,Matone07,Inoue18}, 
but never tested on interferometers so far.
The systematic uncertainties associated with the signal produced by this Newtonian calibrator, or “NCal”,
are different from the usual calibration techniques and might lead to improved absolute calibration errors compared to existing methods.

To investigate this approach, a simple NCal prototype has been built and tested on the Advanced Virgo detector.
The paper reports on this investigation, first presenting the principle of the calibrator in section 2, then describing the prototype in section 3.
The signal observed and the first checks made with the system are presented in section 4,
and the possible noise sources induced by operating the device are discussed in section 5.
Finally the plans and projected systematic uncertainties of this technique are presented in the last section.

%% file: model.tex
\section{Predicting the signal induced by the NCal}

In this section, we derive the effect of a simple rotor on the position of one of the interferometer mirrors, 
first using a very simple point-like model of a device aligned on the beam axis to show 
the basic characteristics of the signal to be observed in the $h(t)$ gravitational wave strain. 
Then we consider an off-axis device and a more realistic, extended rotor. 
Finally we show the results of a finite element model that turns out to give predictions very close to the analytical model.

These computations are done for a rotor made of two masses. 
More complex geometries, based on three masses for instance, could be interesting to study systematic effects, 
but the method to compute the expected signal remains the same.

\subsection{Point-like model of a rotor aligned on the beam axis}

\begin{figure}[h!]
  \begin{center}
    \includegraphics[angle=0,width=0.7\linewidth]{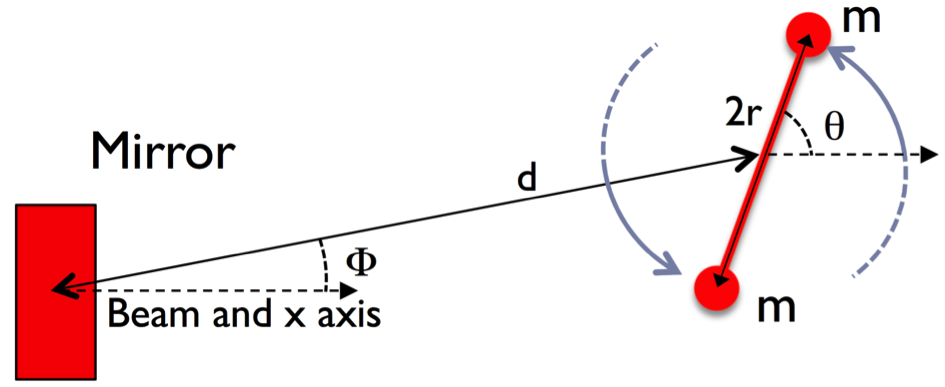}
    \caption{Sketch of a rotor made of two masses next to a mirror}
    \label{fig:sketch}
  \end{center}
\end{figure}

The simplest possible version of an NCal is a system of two masses m separated by a distance $2r$, 
on a rotor rotating at frequency $f$, 
located at a distance $d$ from the mirror of mass $M$ (see figure \ref{fig:sketch}).
			
For this initial computation, we assume that the rotor center is 
aligned on the beam axis that is our x axis ($\Phi = 0$). 
The coordinates of one mass $m$ relative to the mirror position are then $(d + r \cos\theta); (r \sin\theta)$. 
If we assume point-like masses for the actuation masses and the mirror, 
and do a second-order Taylor expansion in $\epsilon=r/d$, 
the gravitational force applied by the first mass on the mirror along the beam axis is:

\begin{equation} \label{eq:F1x}
F_{1x} \approx \frac{GMm}{d^2}\Big[1-\tfrac{3}{2} \epsilon^2 - 2\epsilon \cos\theta 
+ \tfrac{9}{2}\epsilon^2 \cos^2\theta\Big] 
\end{equation}

and similarly for the second rotor mass:

\begin{equation} 
F_{2x} \approx \frac{GMm}{d^2}\Big[1-\tfrac{3}{2} \epsilon^2 + 2\epsilon \cos\theta 
+ \tfrac{9}{2}\epsilon^2 \cos^2\theta\Big] 
\end{equation}

The total force acting on the mirror that depends on the rotor angle is therefore:	 

\begin{equation} 
F = \frac{9GMm}{d^2}\Big[\epsilon^2 \cos^2\theta\Big] 
\end{equation}

or

\begin{equation} 
F = \frac{9}{2}\frac{GMmr^2}{d^4}\Big[\cos(2\theta)\Big] 
\end{equation}

When the masses are rotating at a frequency $f_{rotor}$, 
the time dependent force and therefore the mirror displacement and apparent $h(t)$ signal occur at 
twice this frequency: $f_h = 2f_{rotor}$. 
If this frequency is well above the proper frequency of the mirror suspension (about 0.6 Hz), 
the amplitude of the mirror motion at $f_h$ is:
\begin{equation}
	a(f_h) = \frac{RP}{f_h^2}
\end{equation}
with
\begin{equation}
 	R = \frac{Gmr^2}{8\pi^2}
\end{equation}
a parameter that depends only on the rotor geometry, and 	
\begin{equation}
P = \frac{9}{d^4}
\end{equation}
a parameter that captures the position of the center of mass of the rotor.

It is important to note that this motion is independent of the mirror mass, 
a feature that makes the motion induced by the NCal independent of the mirror details in first approximation.

\subsection{Point-like model of a rotor not aligned on the beam axis}\label{sec:2.2}

In the more general case where the rotor is off axis by an angle $\Phi$ relative to the beam axis 
in the interferometer plane, 
and offset by a distance $z$ below or above this plane.
The NCal rotation plane is assumed to be parallel to the interferometer plane. 
Then the coordinates of the first mass relative to the mirror become:
	 $(d\cos\Phi + r \cos\theta); (d\sin\Phi + r \sin\theta); z$

	where $d$ is still the projection of the rotor-mirror distance in the interferometer plane.
Then the parameter P describing the rotor position becomes:
\begin{equation} 
	P = \Big((\frac{15}{Z} - 6)^2\cos^2\Phi+ 6^2\sin^2\Phi\Big)^{1/2}Z^{-5/2}d^{-4}
\end{equation}
with 
\begin{equation}
Z = \Big(1+\Big(\frac{z}{d}\Big)^2\Big)
\end{equation}
Three remarks related to the systemic uncertainties could be made on this parameter:
\begin{itemize}
\item When $z$ is close to zero, the $Z$ parameter becomes close to 1 and independent of $z$.
This means that if the rotor vertical position is close to the interferometer plane, 
then the parameter $P$ and therefore the mirror displacement only weakly depend on the exact vertical position.
\item The parameter $P$ varies slowly as function of the $\Phi$ angle. 
Therefore this angle is not a critical parameter when computing the strength of the NCal signal. 
\item If $z = d/2$, then $Z = 5/4$ and $P$ becomes independent of the $\Phi$ angle.
\end{itemize}
 
It is interesting to note that even if the NCal is located on the side of the mirror ($\Phi=\pi/2$), 
the signal induced is still two thirds of the maximum amplitude.

\subsection{Extended rotor}

Real rotors are not point-like masses. 
One simple geometry that avoids large air motion is to consider a disk of thickness $b$, 
made of two sectors with an opening angle $\alpha$ and inner and outer radii $r_{min}$ and $r_{max}$,
like in figure \ref{fig:proto}.
Assuming that the rotor size is small compared to the distance to the mirror, by integrating over the radius, 
opening angle and thickness, with $\rho$ being the rotating mass density we can get the overall coupling factor:
\begin{equation} \label{eq:Rfull}
	R = \frac{G b \rho}{8\pi^2}\sin\Big(\frac{\alpha}{4}(r_{max}^4- r_{min}^4)\Big)
\end{equation}
The maximum coupling factor is achieved with $\alpha$ = $\pi/2$, 
i.e. when the rotor is divided in four sectors of the same volume, alternating empty and full volumes. 
In this case, any error on the opening angle of the rotor cancels at first order. 
The coupling factor depends only on the rotor thickness and radii, mostly its outer radius.

\subsection{Correction for an extended mirror}

A finite element model has been developed to include more refined effects 
when the detailed mirror and rotor geometry is taken into account. 
This model uses the Advanced Virgo mirror geometry (35 cm diameter, 20 cm thickness), 
neglecting the "ears" used to hang the suspension fibers on the sides of the mirror, since their mass is very small. 
The grid used divides each of the rotor and mirror in one thousand small parts, a number that was verified be large enough. 
For a given rotor angle, all interactions between these parts are computed using the exact 
Newtonian force formula and summed to get the overall resulting force. 
The amplitude of the force variation, and therefore the NCal signal amplitude, 
is then derived by changing the rotor angle in small steps till the rotor makes half a turn.

\begin{figure}[h!]
  \begin{center}
    \includegraphics[angle=0,width=0.7\linewidth]{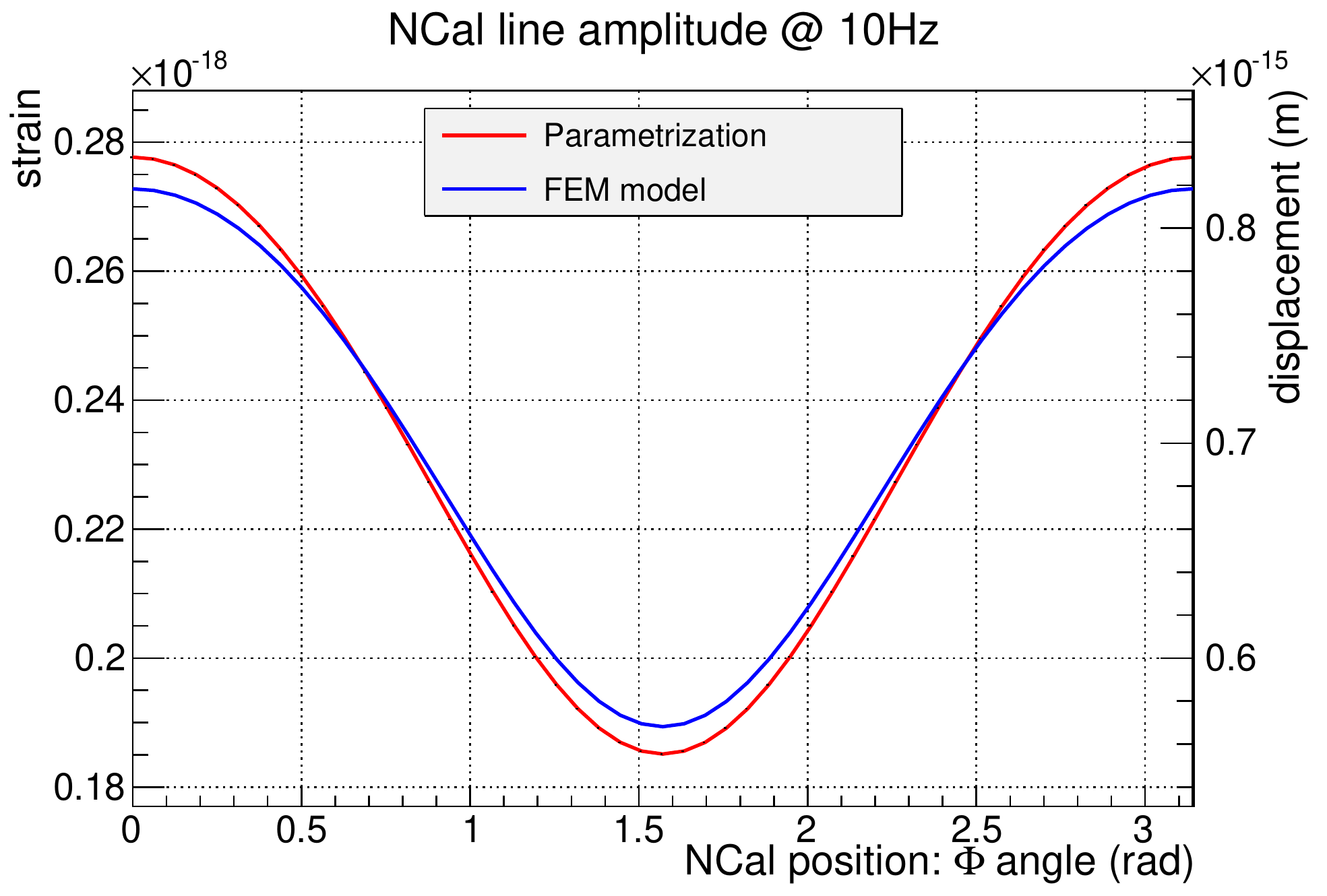}
    \caption{Expected NCal line amplitude at 10 Hz induced by a rotor 
          made of two masses as a function of the NCal position relative to the beam axis. 
	On the left vertical axis, the signal is expressed as strain. 
	On the right vertical axis, the signal is expressed as mirror displacement. 
	The rotor geometry is the one described in section 3. The NCal is at 1.32 m of the mirror, 
        at the same heigth ($z = 0$).}
    \label{fig:model}
  \end{center}
\end{figure}

Figure \ref{fig:model} shows the NCal signal amplitude as a function of the rotor angular offset relative 
to the beam axis (red curve), compared to the result of the simple analytical model 
described in the previous section (blue curve). 
The values used for the mirror position correspond to the configuration of the test presented 
in the following sections ($d = 1.32$~m), except for the vertical position
since in the test, the vertical offset was very close to the magic value of half the distance that 
cancels out the dependence on the $\Phi$ angle (see the third remark in section \ref{sec:2.2}). 
Therefore, for this figure, we set the vertical position to $z = 0$ that is the preferred value 
to minimize the coupling with the vertical position uncertainties.
The two models agree to better than 2.2~\%, 
indicating that systematic effects due to remaining inaccuracies in the model are much smaller. 
For a more distant rotor, the agreement is even better. 
In the case of the configuration used in the next section, the correction 
when using the FEM model instead of the parameterization is 1.4~\%. 

%% file: proto.tex
\section{NCal prototype}
The NCal used for the tests presented in this paper was built as a simple prototype, with limited resources. 
The goal of the tests was to check that a calibrated "Newtonian" signal could be injected as expected, 
without disturbing the operation of the interferometer. 
Therefore, its design is far from being optimized. 

\begin{figure}[h!]
  \begin{center}
    \includegraphics[angle=0,width=0.5\linewidth]{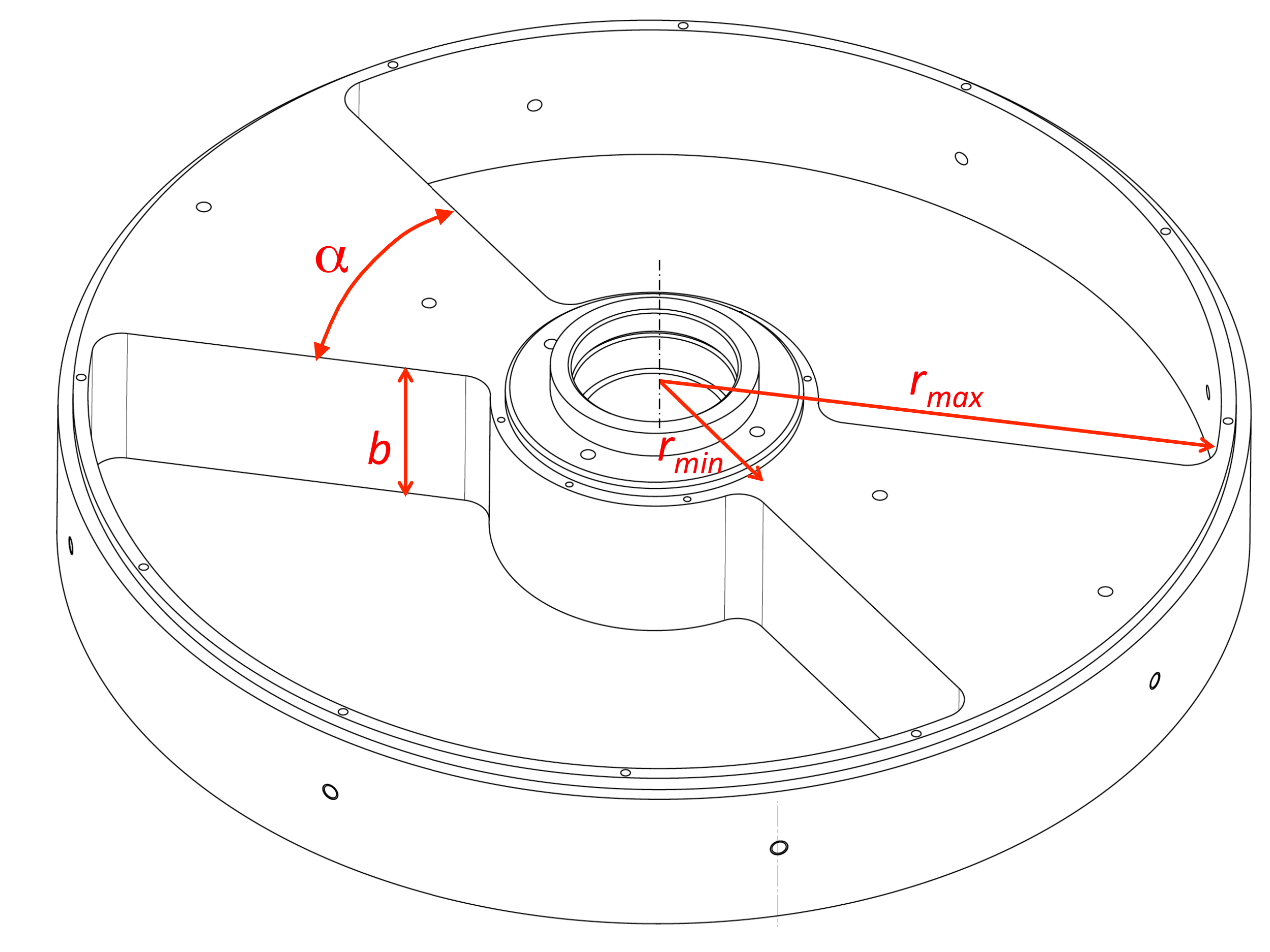}
    \caption{View of the NCal rotor used without its top plate.}
    \label{fig:proto}
  \end{center}
\end{figure}

The NCal rotor was symmetrical around its axis, except for two sectors having an opening angle of 45 degres, 
an inner radius of 55.0 mm and outer radius of 190.0 mm. 
The thickness was 52.0 mm (see figure \ref{fig:proto}). 
Minor non-symmetrical features like a few holes for the screws used to close the rotor 
with a top plate are neglected in this study. 
Although a dense material would be in principle useful, 
the rotor was made of aluminum (7075 T6, density 2.80 kg/liter), 
to reduce the cost and simplify the machining of parts. 
Therefore each sector weighs about 2.06 kg.

Two heavy-duty ball bearings were used to hold the rotor, 
which results in large friction and the need to increase the power of the motor used to spin the rotor. 
This is the reason why the first tests were made with a limited rotation speed. 
During the tests, various commercial DC brushless motors were used. 
For the results presented here, the power needed was a bit more than 100~W when the rotor was spinning at 50 Hz. 
Two pulleys and a toothed belt were used to make the motor turn 1.2 times faster than the rotor
to disentangle vibrations coming from the motor and those coming from the rotor.

In addition to recording the parameters of the motor controller, 
a simple shadow-meter sensor was used to monitor the rotor rotation and absolute position.

The rotor was resting inside an aluminum square box, while a simple cover was put on top of the belt. 
The device was acoustically fairly noisy. 
The NCal was simply laid on the base of the mirror vacuum chamber, but in-air, 
with just one centimeter of foam as insulation. 
Its position was measured with a simple ruler with an accuracy of several millimeters.

%% file: test.tex
\section{Comparing the injected and recovered signals}
\subsection{Test description}
The first test was made at the Virgo site on July 4, 2017. 
This was a very preliminary test since the NCal was only able to spin up to 20 Hz, 
but allowed us to verify that the induced signal was indeed visible, 
even though Advanced Virgo was still in commissioning phase with a sensitivity 
roughly three times worse than that later achieved during O2. 

After installing a more powerful motor, another test was made on November 6, 2017 
with the NCal positioned at three different locations. 
Half an hour of data was collected with the NCal at its closest position to the mirror: 
the NCal was $d = 1.32$~m away from the mirror in the interferometer plane, 
with an angle $\Phi= 0.241$ radian relative to the beam axis, 
and was $z = 0.643$ m below the interferometer plane. 
This means that the expected displacement induced by the NCal 
at frequency $f$ was $3.3\times10^{-14}(1\text{Hz}/f)^2 \text{m}$.

During this test, the rotor frequency was swept from zero to 50 Hz, 
pausing at a few frequencies for one or two minutes. 
The $h(t)$ signal was then reconstructed using the same code and parameters as during the O2 run~\cite{calibVirgo2018}. 

\subsection{Typical SNR of the NCal lines}

\begin{figure}[h!]
  \begin{center}
    \includegraphics[angle=0,width=0.7\linewidth]{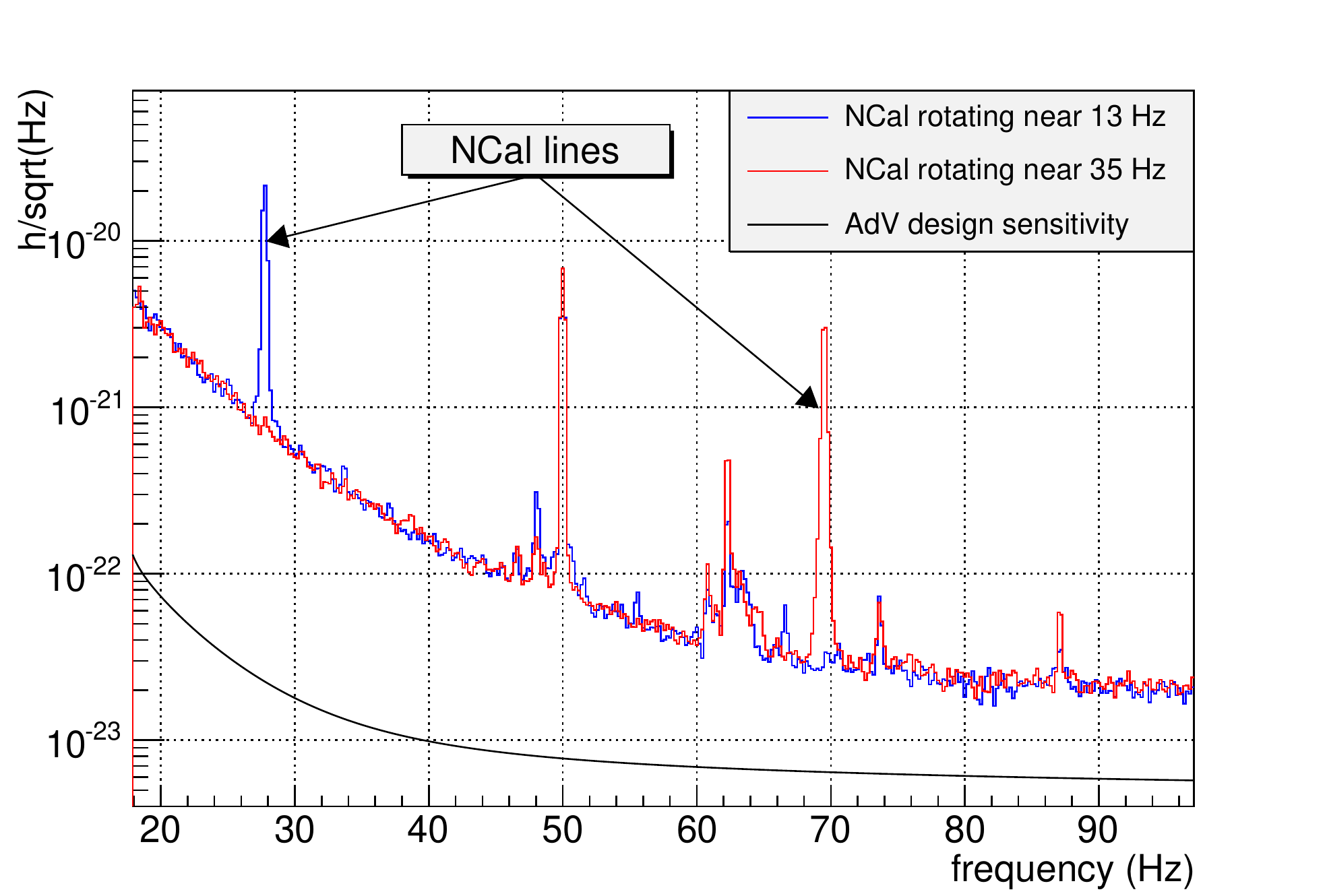}
    \caption{Measured spectrum of the $h(t)$ signal with the NCal rotating at two different speeds. 
           The NCal signal is observed at twice the NCal rotation speed.}
    \label{fig:spectrum}
  \end{center}
\end{figure}

Figure \ref{fig:spectrum} shows the $h(t)$ spectrum of the AdV detector for two such datasets, 
where only the NCal frequency has been changed. 
The only significant difference between the two spectra is the position of the NCal line. 
The signal to noise ratio (SNR) of the NCal line was about 50 for 5 seconds long FFT. 
Further SNR improvements are expected when the detector sensitivity gets 
closer to the predicted AdV design sensitivity.

\subsection{Checking the amplitude and phase of the NCal line}

\begin{figure}[h!]
  \begin{center}
    \includegraphics[angle=0,width=1\linewidth]{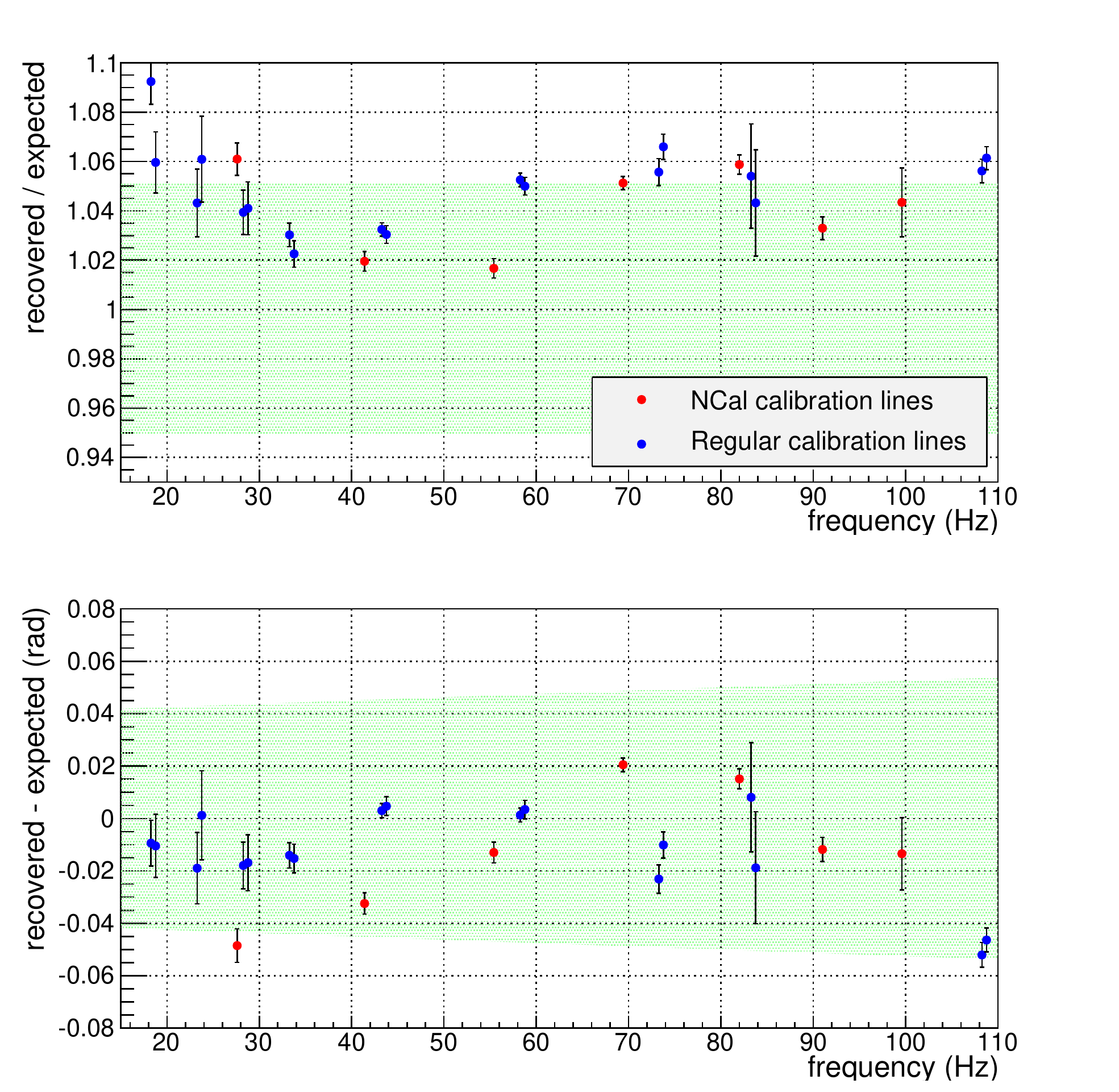}
    \caption{Recovered/expected amplitude ratios and phase differences for NCal and regular coil/magnet calibration lines. The green shaded areas represent the O2 final calibration error budget.}
    \label{fig:ratios}
  \end{center}
\end{figure}

It is possible to do a more quantitative validation with the NCal calibration lines 
by comparing them to their expected values. 
The result is reported in figure \ref{fig:ratios}, where the top plot shows the ratio 
between the recovered and expected  NCal line amplitudes (red dots), 
and the bottom plot shows the phase differences. 
Each point corresponds to about one minute of data. 
On the same day, a set of regular calibration lines injected with the coil-magnet actuators were collected 
and are displayed on the figure (blue dots) to compare both calibration methods. 
The error bars correspond only to statistical uncertainties 
(one standard deviation or 68\% confidence level as for all uncertainties in this paper), 
which are below 1\% for some of the NCal lines, thanks to the high SNR of the NCal signal. 
The green shaded area corresponds to the overall uncertainties of the O2 final calibration error budget~\cite{calibVirgo2018}. 

The recovered/expected amplitude ratios and the phase differences are slightly off the O2 calibration 
uncertainties, although the size of their variations is compatible with those uncertainties. 
The residual offsets are due to a not very accurate calibration of the detector on that day 
because one of the four coils of the north end cavity mirror was not working due to some commissioning issue. 
Since the four coils have the same design and independent driving electronics, 
we expect a 3/4 reduction of the driving force. 
The calibration was therefore simply corrected by a 4/3 factor, 
since the individual coils calibration was not available, 
leading to such possible residual offsets.

However, the regular calibration lines and the NCal lines on the same day show very similar results. 
The same wavy shape has been observed during all O2 calibration studies and 
seems to be a bias of the $h(t)$ reconstruction. 
This confirms that both calibration methods provide compatible results.

\subsection{Checking the NCal position relative to the mirror}\label{sec:4.4}

We made a test by installing the NCal in two different positions, located at the same distance $d = 1.878$~m 
from the mirror, but on the exact opposite side of the mirror. 
In that case, we expect the NCal signal $h(t)$ amplitude to be the same with just a sign flip.

If the mirror position has an offset that is difficult to spot since it is inside a vacuum chamber, 
then the NCal-mirror distance will be smaller on one side and larger on the other side. 
Because the NCal $h(t)$ amplitude varies as the fourth power of the distance, 
the ratio of the observed amplitudes will therefore be 8 times this mirror offset times the distance $d$.

For this measurement we collected about two minutes of data with a 55 Hz line in $h(t)$ at both positions. 
The ratio between the two amplitudes is $R=1.010\pm0.006$, which translates into a mirror offset position 
of $2.3 \pm 1.4$~mm, compatible with the possible mirror to NCal position uncertainty.

Of course, the asymmetry could also come from an error on the NCal positioning 
that was again of the order of a few mm. 
The phase difference, after correcting for the sign flip, has been measured to be $6.2 \pm 6.0$~mrad, 
not a significant deviation from the expected zero value. 

The main result of this test is not the derived offset value, 
but to show that with just a couple of minutes of data, with a sensitivity that was 
an order of magnitude worse than the design sensitivity, 
the consistency check can already be done to better than 1~\%.

\subsection{Checking the impact of the NCal distance on the mirror}

Another test was made during a short commissioning period on February 2, 2018. 
This happened right after the installation of monolithic suspensions for some of the interferometer mirrors. 
There had been no time to recover a good sensitivity and the BNS inspiral range was just below 20 Mpc, 
about two thirds of the Advanced Virgo sensitivity previously achieved during the O2 run. 
The purpose of the test was to collect more data points at different frequencies 
with two different NCal-mirror distances, in order to
verify that the $h(t)$ signal changes as predicted by the model. 
Data were collected with the NCal located at 1.78 m and 2.10 m from the mirror (total distance), 
still off-axis by 0.64 m in the vertical direction. 

\begin{figure}[h!]
  \begin{center}
    \includegraphics[angle=0,width=1.1\linewidth]{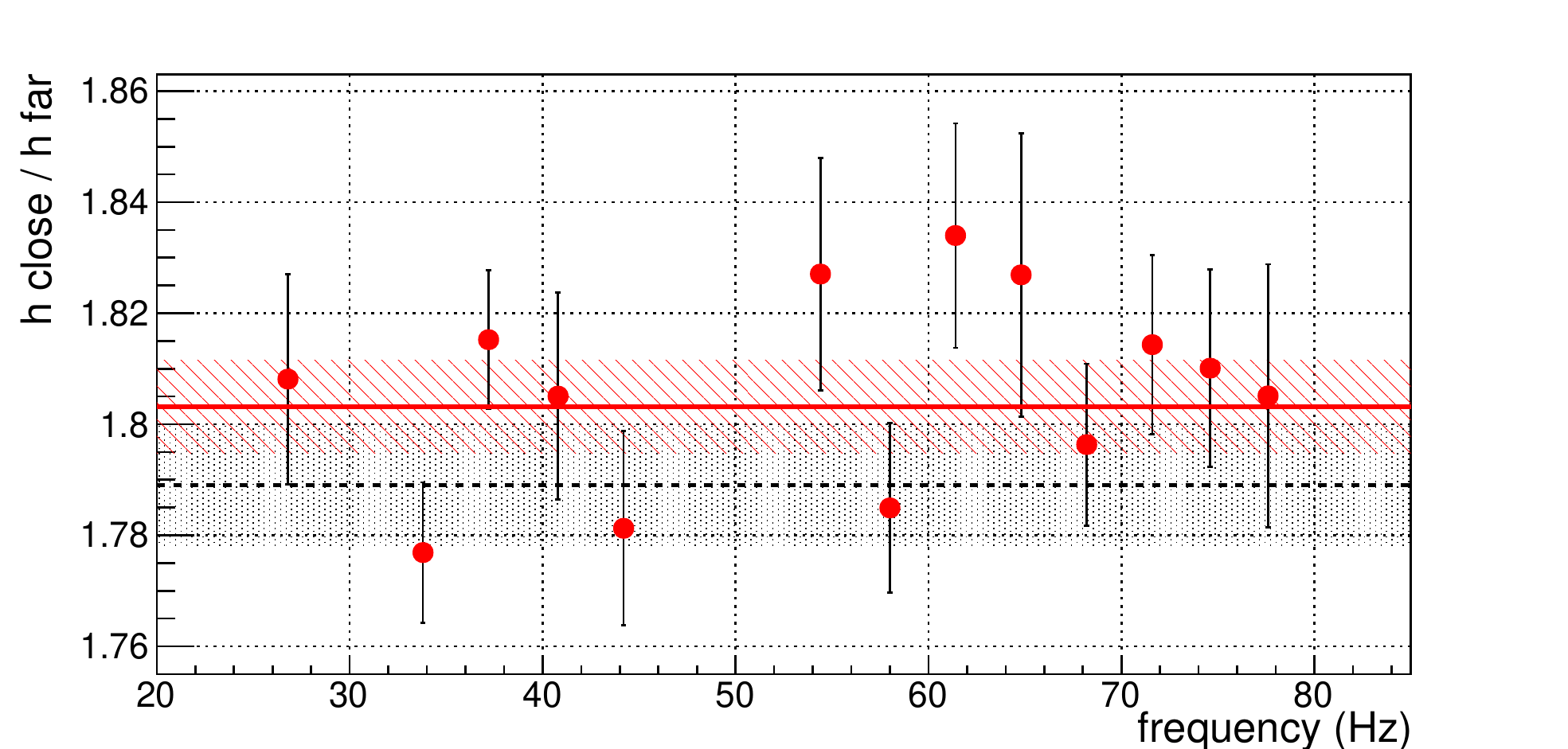}
    \caption{Ratios of the calibration line amplitudes when the NCal is at the position 
       close to the mirror relative to the distant position. 
       The red line is the average value, the black dashed line the expected value.
       The shaded areas are the one standard deviation uncertainty on the average and expected values.}
    \label{fig:ratioDistance}
  \end{center}
\end{figure}

The ratios of the observed calibration line amplitudes in $h(t)$ at 
the two distances are presented in figure \ref{fig:ratioDistance}. 
The average value is $1.803\pm0.008$, in agreement with the expected value of $1.788 \pm 0.011$. 
The error on the measurement is statistical. 
The error on the prediction comes 
from an NCal positioning uncertainty of $\pm$ 2 mm and includes the uncertainty 
on the mirror location in the horizontal plane coming from the measurement described 
in the previous section and a 1 cm uncertainty in the vertical direction. 
If instead of measuring this ratio in the $h(t)$ channel we use the raw interferometer dark fringe 
photodiode channel to avoid possible residual small bias due to the $h(t)$ reconstruction 
(like the wavy shape of figure \ref{fig:ratios}), 
the observed value is $1.790\pm0.012$, again in good agreement. 
This ratio confirms that the NCal signal goes as the fourth power of the NCal distance 
and excludes significant $d^{-2}$ coupling 
that would come for instance from a simple electromagnetic coupling.

%% file: coupling.tex
\section{Looking for non-gravitational coupling}

The injection of the NCal signal relies on the well-established Newton laws. 
It is not surprising to observe such a signal in the $h(t)$ stream. 
However, the NCal can produce noises that might couple to the mirror 
through other mechanisms than the direct gravitational force. 
This could change the strength of the observed signal as well as its phase. 
The good match between the predicted and observed phase is an encouraging indication 
that such possible noises have a small impact.

\begin{figure}[h!]
  \begin{center}
    \includegraphics[angle=0,width=1\linewidth]{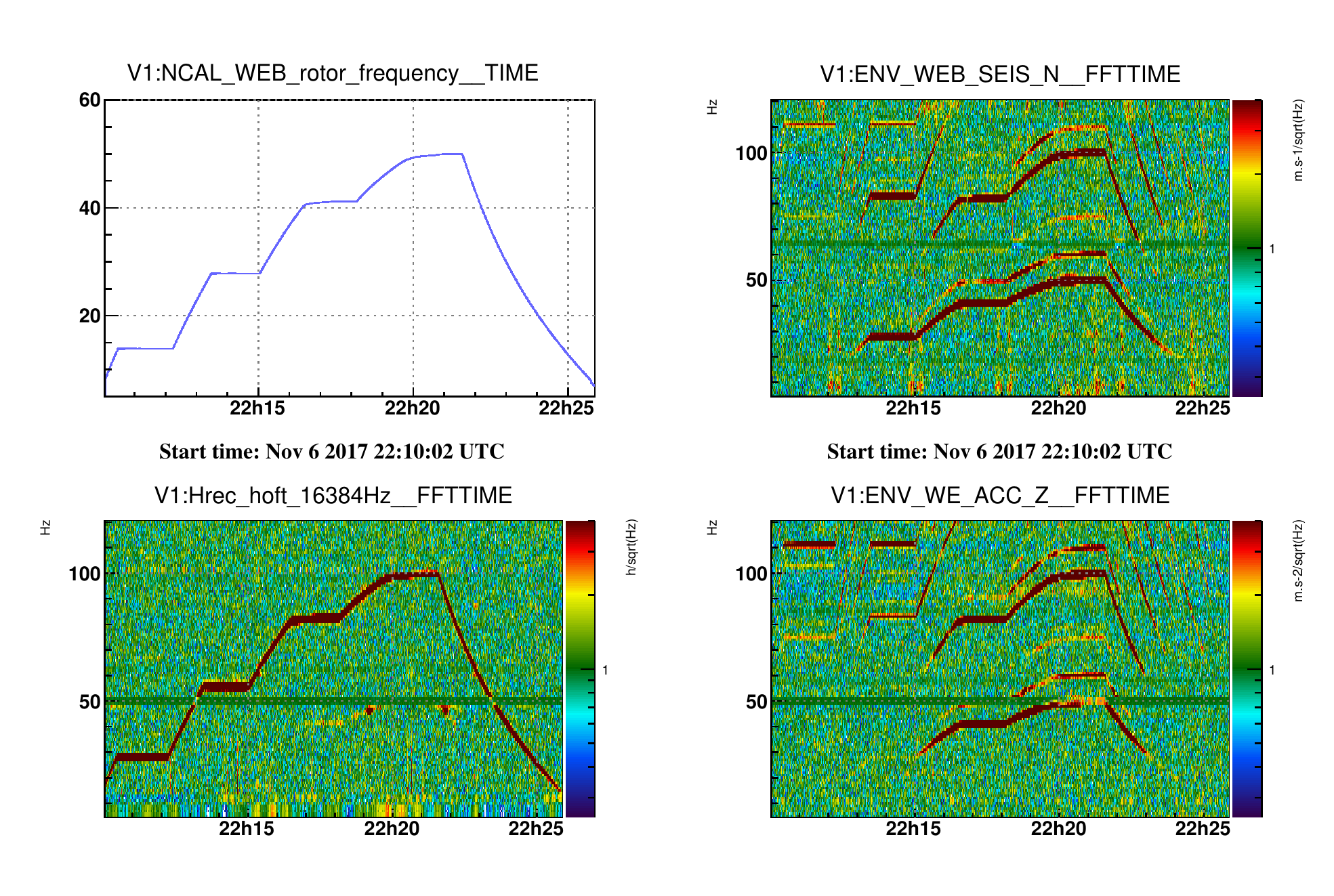}
    \caption{The top left plot is the frequency of the rotor rotation in Hz, 
the bottom left plot is the time-frequency plot for the $h(t)$ channel. 
The top right plot is the time-frequency plot for an accelerometer 
connected to the main vacuum chamber and close to the NCal, 
the bottom right plot for a seismometer channel. 
All time-frequency plots are made with a 1 Hz, 1 second resolution 
and are relative to the median average spectrum.}
    \label{fig:coupling}
  \end{center}
\end{figure}

Various probes are available around the mirror vacuum chamber to monitor the environment. 
As explained earlier, the NCal prototype was not a very quiet device and the seismic probes 
and accelerometers connected to the vacuum chamber recorded well visible signals, 
while the magnetometers did not and the microphone turned out to be less sensitive 
than the accelerometer that was closer to the NCal. 
Figure \ref{fig:coupling} presents the time evolution of the rotor frequency and time-frequency plots for $h(t)$, 
the most sensitive seismometer and accelerometer for fifteen  minutes of measurement. 

The $h(t)$ time-frequency plot is fairly clean, showing mostly the expected NCal signal 
plus some extra noise around 45 Hz when the NCal is spinning at this frequency. 
This is a frequency band where the detector was noisier, 
as can be seen in figure \ref{fig:spectrum}. 
On the other hand, the seismometer and accelerometer show many simultaneous lines at the rotor frequency 
and its harmonics, at the motor frequency that was spinning 1.2 times faster than the rotor, 
and at various combinations of those frequencies. 
Observing all those lines in the environmental probes while $h(t)$ displays only 
the signal at twice the rotor frequency (besides the 45 Hz noise) shows that the non-gravitational 
coupling of the external noises to the $h(t)$ signal is much smaller than the direct gravitational coupling. 
Due to limited time, we could not make a quantitative estimate of how small these non-gravitational couplings are, 
but given the fact that this prototype was noisy, 
we are confident that a better designed NCal system will not degrade the $h(t)$ calibration or sensitivity.

%% file: next.tex
\section{Next steps and possible systematic uncertainties}
 
A new NCal prototype is currently being built at LAPP with various improvements. 
It will be a more compact system in order to reduce the acoustic and therefore seismic noise 
that could be produced. Its diameter will be only 205~mm. 
To compensate for this reduction, the opening angle will be corrected from the $\pi/4$ of the first prototype 
to the optimum value of $\pi/2$ and it will be installed in the interferometer plane. 

Both changes will reduce the associated systematic uncertainties. 
We expect that the dominant ones will be:
\begin{itemize}
\item The thickness of the rotor (80~mm) that could be measured to better than 0.1~mm, leading to a 0.13~\% uncertainty
\item The external diameter of the rotor that could be measured to better than 0.1~mm, leading to a 0.4~\% uncertainty
\item The density of the material used that could be known/measured to better than 0.1~\%,
leading to an uncertainty of the same amount
\item The NCal position that could be measured to better than 1~mm, leading to a 0.3~\% uncertainty
\item The distance to the mirror that could be measured to better than 1~mm using the technique of 
installing the NCal on both sides of the mirror as described in section \ref{sec:4.4},
leading to an uncertainty of less than 0.4~\%.

\end{itemize}

This would give a total uncertainty around 1~\%, without very challenging techniques, 
leaving room for future improvements.

The long term stability of the NCal signal is expected to be good compared, for instance to a PCal with aging effect.  
The dominant effect might by the thermal expansion of the NCal. 
Since the signal goes as the fourth power of the NCal diameter, a variation of $\pm 1$ degree 
(more than the typical temperature variation in the Virgo buildings) 
for a rotor made of aluminum (worst case) will induce a 0.1~\% variation of the NCal signal, 
below the expected systematic uncertainties.

%% file: conclusion.tex
\section{Conclusions}
 
The old idea of calibrating a gravitational wave detector using the varying gravitational field 
of moving masses has been tested for the first time on an interferometer. 
The observed amplitude and phase of the NCal calibration lines match very well those of the 
usual coil-magnet calibration. 
The observed signal dependence on the distance between the NCal and the mirror is in good agreement 
with the expected fourth power law. 
Good control of the mirror position, inside the vacuum chamber, 
relative to the outside NCal position has been demonstrated using 
a set of two NCal positions located on both sides of the mirror. 
Although this first prototype was fairly noisy, 
no strong influence produced by non-gravitational coupling has been observed. 

All these results are convincing arguments to further develop the NCal technique 
that might become the prime reference for absolute $h(t)$ calibration. 
A new prototype is currently being built and tested. 
Its absolute uncertainty is expected to be below one percent. 
The frequency range of the NCal that was limited to one hundred Hz with the first prototype 
is expected to double with this second prototype, covering the most sensitive part 
of the detector bandwidth that is critical for parameter estimation of coalescing binary systems. 
Future developments could extend this frequency range, 
but in any case this will be already enough to cross calibrate the photon calibrators 
and coil-magnet actuators that have a much larger bandwidth.

NCal devices should have a very good long-term stability. 
Improving them could bring the absolute $h(t)$ calibration well below the one percent expected 
for the second prototype. 
They should be easily moveable from one interferometer to another, 
providing a very accurate calibration of the network of gravitational wave detectors, 
enabling high precision measurements and the associated science.